\documentclass[11pt]{article}

\usepackage[utf8]{inputenc}
\usepackage[T1,tone,safe]{tipa}

\usepackage{latexsym,amsmath,amssymb,theorem,epsfig}

\usepackage{cite}

\usepackage{pifont}
\usepackage{array}

\usepackage{setspace}
\usepackage{fullpage}
\usepackage{url}
\usepackage{multirow}
\usepackage[colorlinks=true,linkcolor=blue,urlcolor=blue,citecolor=blue]{hyperref} 

\usepackage{mciteplus} 
\mciteSetMidEndSepPunct{\nolinebreak;\space}{.}{\relax} 

\DeclareFontFamily{OT1}{rsfs10}{}
\DeclareFontShape{OT1}{rsfs10}{m}{n}{ <-> rsfs10 }{}
\DeclareMathAlphabet{\mathscript}{OT1}{rsfs10}{m}{n}

\numberwithin{equation}{section}

\newcommand{\ns}{\normalsize}

\def\g{\gamma}

\def\s{\sigma}

\def\gsim{ \lower .75ex \hbox{$\sim$} \llap{\raise .27ex \hbox{$>$}} }
\def\lsim{ \lower .75ex \hbox{$\sim$} \llap{\raise .27ex \hbox{$<$}} }
\def\be{\begin{equation}}
\def\ee{\end{equation}}
\def\bea{\begin{eqnarray}}
\def\eea{\end{eqnarray}}

\def \td {\tilde}

\def \ha {{1 \ov 2}}

\def\ov{\over}
\def \ci {\cite}

\def \foot {\footnote}

\def\la{\label}\def\foot{\footnote}\newcommand{\rf}[1]{(\ref{#1})}

\def \no {\nonumber}

\def \ed   {\end{document}}

\def \g {\gamma}

\theoremstyle{plain}

{\theorembodyfont{\rmfamily} }

\def \ed {\end{document}}

\newcommand{\SU}{\operatorname{SU}}
\newcommand{\dif}{\mathrm{d}}
\newcommand{\SO}{\operatorname{SO}}
\newcommand{\sn}{\operatorname{sn}}
\newcommand{\SL}{\operatorname{SL}}
\newcommand{\R}{\mathbb{R}}
\newcommand{\GL}{\operatorname{GL}}
\newcommand{\C}{\mathbb{C}}
\newcommand{\diag}{\operatorname{diag}}
\newcommand{\lagr}{\mathcal{L}}

\newcommand{\cmark}{\ding{51}}
\newcommand{\xmark}{\ding{55}}

\def\var{\eta}
\def\varn{\xi}

\def\tM{M}
\def\tN{N}

\def\bC{Q}

\usepackage[normalem]{ulem}

\newenvironment{definition}[1][Definition]{\begin{trivlist}
\item[\hskip \labelsep {\bfseries #1}]}{\end{trivlist}}

\begin{document}

\overfullrule=0pt
\parskip=2pt
\parindent=12pt
\headheight=0in \headsep=0in \topmargin=0in \oddsidemargin=0in

\vspace{ -3cm}
\thispagestyle{empty}
\vspace{-1cm}


\begin{titlepage}

\vspace{-5cm}


\vspace{-5cm}

\title{
  \hfill{\small Imperial-TP-AS-2012-01  }  \\[1em]
   {\LARGE 
   On (non)integrability of classical strings  in  $p$-brane backgrounds
   }
\\[1em] }
\author{
   A. Stepanchuk\footnote{a.stepanchuk11@imperial.ac.uk} \ \  and 
     A.A. Tseytlin\footnote{Also at Lebedev  Institute, Moscow. 
   tseytlin@imperial.ac.uk }
     \\[0.5em]
   {\ns The Blackett Laboratory, Imperial College, London SW7 2AZ, U.K.}} 

\date{}

\maketitle

\begin{abstract}
We  investigate the question of   possible integrability of  classical
string motion     in curved p-brane backgrounds. For example, the
D3-brane   metric   interpolates between   the flat   and the  $AdS_5 \times
 S^5$    regions in which string propagation  is integrable. We find that  while the point-like
string  (geodesic) equations are  integrable,  the equations
describing   an extended string in  the
complete  D3-brane  geometry  are not.  The same conclusion   is
reached for  similar  brane  intersection backgrounds  interpolating
between flat space
and  $AdS_k\times S^k$.  We consider, in particular,  the  case of  the
NS 5-brane -  fundamental string background.  To demonstrate
non-integrability  we  make  a   special  ``pulsating string'' ansatz
for which the  string equations reduce to an effective one-dimensional
system.  Expanding   near  this   simple solution
leads to a linear  differential equation for small fluctuations   that
 cannot be solved in quadratures,   implying    non-integrability  of
the original  set of string equations.
\end{abstract}

\thispagestyle{empty}
\end{titlepage}
\section{Introduction}
Strings in curved space are described  by 2d  sigma models  with  complicated non-linear   equations of motion. 
This precludes,  in most cases, {a}{} detailed  understanding of their dynamics. Integrability selects a subclass of 
models  or target space backgrounds   for which one may hope to  come as close as possible to 
 the situation in flat space, i.e.  to have a  complete quantitative  description 
of classical string motions and  then also  of  the  quantum string spectrum. 

A prominent   maximally symmetric 10-dimensional example  is
provided   by  the superstring in the $AdS_5 \times S^5$  space-time:  
much progress  towards its solution, that  was   achieved 
recently   \ci{bb1,*bb2,*bb3}, is based on its integrability. 
One may wonder   if  integrability  can also apply to string motion in closely  related   
 but less   symmetric  $p$-brane backgrounds \ci{Gibbons:1987ps,ns,db1}, e.g.
   the D3-brane   background \ci{db1,db2,*db3}   which is a  1-parameter (D3-brane charge $Q$) 
 interpolation  between   flat space  ($Q=0$)  and the $AdS_5 \times S^5$ space ($Q=\infty$). 
 While string theory is   integrable   in these two   limiting   cases,  what happens at finite $Q$ 
    is a priori an open question. This is the  question  we are going to address in this paper.

    Since         quantum integrability  of a bosonic model  or  integrability of  its superstring counterpart  is essentially implied 
(leaving aside  some exceptional   cases  or anomalies)   by the 
 integrability of the corresponding   classical bosonic  sigma model 
 the study of the latter is  the first step. 
  Following  the same   method as used previously in \ci{pan,*pan1,za}  we will   demonstrate non-integrability 
    of classical extended string motion  in the D3-brane background  (particle, i.e. geodesic, motion is still integrable). 
    We will also   reach similar conclusions   for  other $p$-brane   backgrounds 
    that interpolate between   flat space and integrable  $AdS_n \times S^k$  backgrounds.

As there is no general classification of  integrable 2d sigma  models,  
let us start with a brief survey   of some known classically integrable  bosonic   models
describing string propagation in curved backgrounds. 
 A  major class of  such models have a target space metric $G_{\tM\tN}$ of  a symmetric space,  including    spheres \ci{pol} 
 (and other related constant curvature spaces \ci{bar,*bar1,*bar2}),  group spaces  (i.e. principal chiral model)   \ci{mz}
 and  various other $G/H$   coset models \ci{mz,pole,*pole1,ei,*ei1,*ei2,*ei3,*ei4}.  Integrability is also   preserved by some  anisotropic 
 deformations  of   the group space   \ci{Che,kl}, the 2-sphere  \ci{fat} and the 3-sphere  \ci{fatt,jg,luk}  models.
 Attempts to find ``non-diagonal'' generalisations of the principal  chiral model  by studying 
  conditions   for  the existence of the  corresponding 
  zero curvature (Lax) representation  were made in \ci{S,oh} and also in \ci{Moh}. 

 Allowing  for the parity-odd   antisymmetric tensor  ($B_{\tM\tN}$) coupling  leads, in particular,  to 
 the  WZW  and related gauged WZW models \ci{w,*w1,gw,*gw1,*gw2,*gw3,*cb,*gw4,*gw5,*Nappi:1993ie}. 
 The    generalised   principal chiral  model with the WZ term having an 
  arbitrary coefficient   is also integrable \ci{fadd,*Veselov1,ev,bl}.
  Some  other ``anisotropic''  integrable  models with 3d target space  
   were constructed in \ci{luk}
     and  new examples of integrable coset-type  models   with extra WZ coupling   were found in \ci{bl}.
  
  There are also examples of integrable pp-wave models related to (massive)  integrable 2d models in light-cone gauge,  
 see e.g. \ci{Amati:1988sa,Tseytlin:1992pq,Russo:1994cv,Tseytlin:1995fh,Maldacena:2002fy,*Russo:2002qj,*Bakas:2002kt,*Tirziu:2003wa}.
  
Given an integrable   model   one can generate a family of  other  integrable   models  (i.e. 
find new ``integrable'' target space backgrounds) 
 by 
performing   transformations that  preserve the classical 2d equations of motion, e.g.
using 2d abelian duality (T-duality) or  
 non-abelian duality,   combined  with  field redefinitions (coordinate transformations). 
 Examples include, in particular,  various  gaugings or marginal 
 deformations  of WZW  and related models,  see e.g.
 \ci{pal,*pal1,*pal2,hs,*hs1,wp,hs2,hs3,*hs4,*hs5,*hs6,*hs7,*hs8}; 
  for a more recent  example related to $AdS_5\times S^5$ 
  see  \cite{ml,*ml1,*ml2}.\foot{One may also  find   new 
  models by taking orbifolds  of known integrable models (see, e.g., \ci{d}),
 though division over discrete subgroups   may also lead to a breakdown of integrability (cf. \ci{bv}).} 
 For   discussions of  an   interplay between T-duality and   integrability  see \ci{ri,*ri1,Moh}.

\

Given a  particular string sigma model, to prove its integrability one is   to find a zero curvature representation 
implying its equations of motion. There is no general method for accomplishing this
and the  full set of  necessary  conditions  for  the existence of a Lax pair  is  not known. For example, the presence  of a
non-abelian isometry group  is  not 
 required.\foot{Indeed, generic sigma models   obtained from gauged WZW models by solving for the  2d gauge field  do not have  isometries  (or have only abelian isometries)   and yet, following from a gauged  WZW action, they must be  integrable.} 
 Among the necessary conditions    should,  of course,  be that  any  consistent   truncation
 of the 2d  equations   of motion to a 1d  set of equations   has to
 represent an integrable  
  mechanical system. For example,  rigid string motion  on a sphere  or AdS space 
  are described by an  integrable Neumann-Rosochatius  model \ci{ar,*ar1}. 
  
 In particular, the point-like string, i.e. geodesic, motion  should   be integrable. 
The  geodesic motion in a general  class of higher-dimensional (rotating) black hole backgrounds
  is known  to be integrable   \cite{Page:2006ka,*Krtous:2007xf}.\foot{Also,  a particular  stationary string motion in
   AdS-Kerr-NUT  backgrounds  is integrable  as well \cite{Kubiznak:2007ca}.}
 It is thus  not surprising  that    geodesics  are also   integrable in  $p$-brane  backgrounds  that are   discussed below.\foot{In addition,  
 if the general string motion is integrable, then 
 motion of a 
 subclass of strings with large momentum along
a null geodesic or, equivalently,   string motion in  a  pp-wave background which is a 
 Penrose limit of a given  curved background 
  should of course  be integrable as well. 
  For  the $p$-brane  backgrounds that we are going to consider 
  string motion in the pp-wave   backgrounds representing Penrose limits
   turn out to be solvable  (see e.g.
\cite{Blau:2002dy,Berenstein:2002jq,*Metsaev:2001bj,Fuji:2002vs,*Ryang:2002fj,Papadopoulos:2002bg,*Blau:2003rt})
as in the light-cone gauge description   the string moves   in a quadratic potential.
However,  since   the coefficient of this potential   happens   to depend
 on $\tau$ in  general, the 
  corresponding models  do   not appear   to be   integrable.
}

To disprove integrability of a given sigma model 
  it is  therefore   sufficient   to   show  that there exists at least   one 1d truncation   of  its equations of motion 
  that is not  an integrable  system  of ordinary second-order differential equations. 
  This can be  done by   demonstrating that the corresponding  motion is chaotic, i.e. 
  by  showing that  the  equations for  small variations around solutions in phase space
  cannot be solved in  quadratures
  \cite{MoralesRuizRamis2001}.     We  shall 
  review this  method  in  section \ref{sec:ani}  (and Appendix  \ref{sec:dgt})  and  illustrate it on  a  simple example
   of the non-integrable two-particle system with the potential $V(x_1,x_2)=\frac{1}{2}x^2_1x^2_2$.

 This     approach  was recently used  to  show that  string motion in 
  $AdS_5\times T^{p,q}$, $AdS_5\times Y^{p,q}$ \cite{pan,za}
   and  ``confining'' supergravity backgrounds \cite{Basu:2012ae}  is not integrable.\foot{Spaces like 
    $T^{1,1}$ and $Y^{p,q}$  are special  in that they have  isometries,  
       preserve supersymmetry  and have  integrable geodesics (in fact, super-integrable -- 
       with more conserved 
  quantities than degrees of freedom  \cite{cel,*cel1}). Yet, the corresponding 2d sigma models  fail to be integrable.}
         Here  we shall  follow   a  similar method    to study  integrability of strings   in  curved  backgrounds   that 
   interpolate between two  integrable  limits -- flat space and  $AdS_{n}\times S^m\times T^k$.
   
   The main  cases include the D3-brane, the D5-D1  background   and the four  D3-brane intersection 
   that interpolate between  flat space and $AdS_5 \times S^5$, $AdS_3 \times S^3 \times T^4$ and $AdS_2 \times S^2 \times T^6$
   respectively. While geodesic motion in these  backgrounds is integrable, making a special ansatz for string motion 
   we will find that the  resulting 1d system of equations is not integrable by applying the variational 
   non-integrability technique for Hamiltonian systems  as used  in  \cite{za}  (section \ref{sec:pbranes}).
   
   In section  \ref{sec:ns5f1}  we shall  study the  NS5-F1   background  and show that  integrability is absent for   generic values of the two charges $Q_1$ and $Q_5$,  but 
     it is restored in the limit  when  $Q_5 \to \infty$.

   In Appendix \ref{sec:dgt} we also give a brief summary of the ideas from
 differential Galois theory on which the non-integrability techniques for Hamiltonian systems are based on.
    In Appendix \ref{sec:ppWave}   we shall   demonstrate non-integrability of string propagation  in the
     pp-wave  geometry T-dual to the 
 fundamental string background  \cite{fs1,Tseytlin:1995fh}.


\section{\label{sec:ani} On the non-integrability of  classical Hamiltonian systems}

Integrability of classical Hamiltonian systems is closely related to the behaviour of variations around phase space curves. Based on this observation,   
ref.\cite{ziglin1982} obtained  necessary conditions for the existence of additional functionally independent integrals of motion. These conditions 
are given in terms of the monodromy group properties of the equations for small variations around phase space curves. 
Using differential Galois theory, refs.\cite{Morales1994140,MoralesRuiz2007845,MoralesRuizRamis2001} improved these results by showing that integrability implies that the identity component of the differential Galois group of variational equations
 normal to an integrable plane of solutions must be abelian (see  Appendix A for some  details).
 
This necessary integrability condition allows one to show non-integrability from the properties of the Galois group. 
However,  determining the Galois group can be difficult and therefore 
one usually takes a slightly different route in analysing the normal variational equation (NVE).

One considers a special class of solutions of the NVE which are functions of exponentials, logarithms,
 algebraic expressions
  of the  independent variables and their integrals and which are known as Liouvillian solutions \cite{2009SJADS...8..279A}. 
 The existence of such solutions is equivalent to the  condition that the identity component $G^0$ of the Galois group is  solvable (see, for example,
  theorem 25 in \cite{Kolchin1948}).
       This in turn implies that if the NVE does not admit Liouvillian solutions then $G^0$ is non-solvable  
 (and thus non-abelian)
   leading to the conclusion of non-integrability.
   
    For Hamiltonian systems the NVE is a second order linear homogeneous differential equation
    and for such equations with rational coefficients Liouvillian solutions can be determined by 
    the Kovacic algorithm \cite{Kovacic86analgorithm} which fails if and only if no such solutions exist. 
    Therefore,  whenever no Liouvillian solutions of the NVE are found by the Kovacic algorithm 
    one can deduce non-integrability of  a Hamiltonian system.
    
    An important subtlety here is that we first need to
     algebrize the NVE, i.e. rewrite it as a differential equation with rational coefficients. 
     This can be done by means of a Hamiltonian change of the independent variable and it was shown in \cite{2006math.ph..10010A} that the identity component of the Galois group is preserved under this procedure.

In summary,  if the algebrized NVE is not solvable in terms of Liouvillian functions or, equivalently, 
if the component of the quadratic fluctuation operator normal to an integrable subsystem does not admit Liouvillian zero modes, the original system is non-integrable. This is consistent with the usual definition of integrability in the sense of Liouville which,
 for Hamiltonian systems, 
  implies that the equations of motion  should be  solvable in  quadratures, i.e. in terms of Liouvillian functions.

The main steps to prove non-integrability of a Hamiltonian system are thus:

$\bullet $ \  Choose an invariant plane of solutions.

$\bullet $ \  Obtain the NVEs, i.e. the variational equations normal to the invariant plane.

$\bullet $ \  Check that the algebrized NVEs have no Liouvillian solutions using the Kovacic algorithm.

This approach was recently used  in  \cite{za,Basu:2012ae} to  show 
 the non-integrability of string motion in several 
 curved  backgrounds and we shall also follow it here.

\

Let us first  explain what  the algebraization procedure of the last step  means. 
A   generic NVE is a second order differential equation of the form
\begin{align}
\ddot{\var} + q(t)\dot{\var} + r(t)\var &= 0 \ . \la{1} 
\end{align}
A change of the  variable  $t\rightarrow x(t)$ is  called Hamiltonian
if  $x(t)$ is a solution of the Hamiltonian system $H(p,x)=\frac{p^2}{2}+V(x)$ where $V(x)$ is some potential. Then  $x$ satisfies the
 first integral equation $\frac{1}{2}\dot{x}^2+V(x) = h$, implying that $\dot{x}$ (and also $\ddot{x}$)   is a function of $x$ only.
Changing the variable $t \to x(t)$  in \rf{1} we obtain (here $'\equiv{\dif\ov \dif x}$)
\begin{align}
\var'' +\Big(\frac{\ddot{x}}{\dot{x}^2}+\frac{q(t(x))}{\dot{x}}\Big)\,\var'+\frac{r(t(x))}{\dot{x}^2}\var &= 0\label{algebrizedEq}\ .
\end{align}
We should now require that the coefficients in this equation are rational functions of $x$.

Let us consider   the following  simple  example  of a Hamiltonian  system  with 
the  canonical variables $(x_i,p_i)$  ($i=1,2$) \cite{Ruiz:1999}
\begin{align}
H &= \frac{1}{2}(p_1^2+p_2^2)+\frac{1}{2}x_1^2x_2^2 . 
\la{2}
\end{align}
A choice of  an    invariant plane,   referred to above,    is 
$P=\{(x_1,x_2,p_1,p_2): \, x_2=p_2=0\}$ along which the solution curves are $x_1(\tau)=\kappa   \tau + \mathrm{const}, \   p_1= \kappa, \  
  \kappa=  \sqrt{2H}$.  The system along  the plane $P$ with the coordinates $(x_1,p_1)$ is
   integrable since the only required constant of motion is provided by the Hamiltonian.

To determine the direction normal to the solution curves of this integrable subsystem we can use the Hamiltonian vector field
 along the curves
 $X^{x_1} = \kappa,\  X^{p_1} = 0,\  X^{x_2} = 0,\  X^{p_2} = 0$.
Thus the normal direction is along $x_2$ and $p_2$ and expanding around $x_2=0$ with $ \delta x_2\equiv \var$ we obtain the 
following normal variational equation (NVE)
\begin{align}
\var''(\tau) = -\bar{x}_1^2(\tau)\, \var(\tau),\ \  \qquad \bar{x}_1(\tau) \equiv  \kappa\tau \ , 
\end{align}
which is solved by the parabolic cylinder functions $D_{-1/2}((i-1)\kappa\tau)$ and $D_{-1/2}((i+1)\kappa\tau)$. 
These are not Liouvillian functions and this  implies the  {\it non-integrability}
 of the original Hamiltonian \rf{2}.

To illustrate the previously discussed algebraization procedure  let us consider the Hamiltonian
\begin{align}
 H = \frac{1}{2}(p_1^2+p_2^2)+\frac{1}{8}(x_1^2-x_2^2)^2  \la{3}
\end{align}
which is equivalent to \rf{2}   by the trivial  canonical  transformation
\begin{align}
 x_1 \rightarrow {x_1-x_2\ov \sqrt{2}} ,\quad  x_2 \rightarrow {x_1+x_2\ov  \sqrt{2}} , \quad 
 p_1 \rightarrow {p_1-p_2 \ov \sqrt{2}},\quad  p_2 \rightarrow {p_1+p_2\ov  \sqrt{2}}.
\end{align}
Restricting to  the invariant plane $P=\{(x_1,x_2,p_1,p_2): \, x_2=p_2=0\}$ we obtain an  integrable subsystem
\begin{align}
\dot{x}_1 = p_1,\qquad \dot{p}_1 = -\frac{1}{2}x_1^3\label{gammaSoln}.
\end{align}
The Hamiltonian vector field on this solution plane is
$X^{p_1} = -\frac{1}{2}\bar{x}_1^3,\  X^{p_2} = 0,\  X^{x_1} = \dot{\bar{x}}_1,\  X^{x_2} = 0$.
The NVE is now obtained by expanding the Hamiltonian equations for the coordinates normal to the invariant plane, i.e. $x_2$ and $p_2$, along \eqref{gammaSoln}. Denoting $\delta x_2 = \var$ we thus obtain 
\begin{align}
\ddot{\var} = \frac{1}{2}\bar{x}^2_1(t)\, \var\ , \qquad\bar{x}_1(t) = 2^{3/4}h^{1/4}\sn\Big(\Big(\frac{h}{2}\Big)^{1/4}t,-1\Big)
\end{align}
where $\bar{x}_1$ is a solution of \eqref{gammaSoln} representing the curves in $P$ parametrised by
 the Hamiltonian $h>0$. We can algebrize this NVE by changing the  variable  $t\rightarrow x=\bar{x}_1(t)$. This leads to  ($'= { d \ov dx}$) 
\begin{align}
\var''-\frac{2x^3}{8h-x^4}\var'-\frac{2x^2}{8h-x^4}\var &= 0\ .
\end{align}
 Further changing   the variable $\var \to \varn = (8h-x^4)^{1/4}\var$ one obtains the NVE in the normal form
\begin{align}
\varn''+\frac{8hx^2+2x^6}{(8h-x^4)^2}\, \varn &= 0.
\end{align}
This equation is solved in terms of hypergeometric functions, i.e. 
it has no Liouvillian solutions. This   implies  non-integrability of \rf{3},    
in agreement with the previous discussion of of the equivalent system \rf{2}.

To summarise,  proving   non-integrability of a Hamiltonian system can be achieved   by 
demonstrating  that  given  an integrable subsystem defined by some invariant 
plane in the phase space,  the  corresponding  normal  variational equations
 are not integrable in    quadratures and thus do not admit 
 sufficiently many conserved quantities.

\def \no {\nonumber}

\def \fo {{1\ov 4}} 

\section{\label{sec:pbranes}String motion in $p$-brane backgrounds}


In what follows we shall consider classical bosonic string motion in a  curved background. In the conformal gauge the action is 
\begin{align}
I &= -\frac{1}{4\pi\alpha'}\int\dif\sigma\dif\tau\,\, \Big[   G_{\tM\tN}(Y)\partial_aY^\tM\partial^aY^\tN +  
\epsilon^{ab} B_{\tM\tN}(Y)\partial_aY^\tM\partial_b Y^\tN \Big]  \ ,  \la{gb}
\end{align}
and the corresponding equations of motion should be supplemented by the Virasoro constraints
\begin{align}
G_{\tM\tN}\,\dot{Y}^\tM\acute{Y}^\tN &= 0\ ,\label{YdotSigmaVirasoro}\\
G_{\tM\tN}(\dot{Y}^\tM\dot{Y}^\tN+\acute{Y}^\tM\acute{Y}^\tN) &= 0\ .\label{YsquaredVir}
\end{align}
In this section we shall consider the case of $B_{\tM\tN}=0$ and the target space metric given by the following $p$-brane ansatz ($Q \geq 0$) 
\begin{align}
 \dif s^2 &= f^{-2}(r)\,\dif x^\mu\dif x^\nu\eta_{\mu\nu}+f^2(r)(\dif r^2+r^2\,\dif\Omega_k^2)\ , \label{Dplike}\\
f(r) &= \Big(1+\frac{\bC}{r^n}\Big)^m,\qquad \ \ \ \ \ n,m\ne 0\label{fofr}\\
\eta_{\mu\nu} &= \diag(-1,+1,...,+1),\qquad\ \ \ \ \   \mu,\nu=0,...,p \no
\end{align}
where $\dif\Omega_k^2$ is the metric on a $k$-sphere.

 For $k=8-p$, $n=7-p\, >0$  
 and $m=\fo$ this  metric describes   the  standard  single $p$-brane  geometry
 in $D=10$   supergravity. It may be supported by a   R-R  field strength  and dilation backgrounds, but since they do not couple to 
the classical  bosonic string we shall ignore them here. 

Keeping  $k,n,m,p$  generic also allows one   to describe lower-dimensional  backgrounds representing  (up to  a   flat factor) 
some special   brane intersection geometries. 
  In particular,  the metric   \rf{Dplike}  with  $nm=1$
  describes a    one-parameter   interpolation  \ci{Gibbons:1993sv,Boonstra:1997dy,*Boonstra:1998yu}   between 
 flat space (for $\bC=0$)    and the $ AdS_{p+2}\times S^k$  space (for  $\bC\rightarrow\infty$).\foot{More precisely, 
 the metric \rf{Dplike} interpolates between the flat-space region ($r^n \gg Q$)  and the 
 $ AdS_{p+2}\times S^k$  region ($r^n \ll Q$).  Moreover,  since $f(r) = \big(c +\frac{\bC}{r^n}\big)^m$
  gives a solution for any constant $c$ including $c=0$,   $ AdS_{p+2}\times S^k$  is 
 also   an exact solution.} 
 
 Special cases include: 
 
  (i) $n=4,\, m=\fo,  \, k=5,\, p=3$:  \, D3-brane    interpolating  between   flat space 
 and $AdS_5 \times S^5$; 
 
 (ii)  $n=2,\, m=\ha,  \, k=3, \, p=1$: \,   D5-D1 (or NS5-F1 with   non-zero $B_{MN}$, see next section)  
   background  \ci{Cvetic:1995bj,Strominger:1996sh} 
 with $Q_5=Q_1=Q$  interpolating  between   flat 10d space 
 and $AdS_3 \times S^3\times T^4$; 
 
(iii) $n=1,\, m=1,  \, k=2, \, p=0$: \,  four  equal-charge D3-brane intersection \ci{Klebanov:1996mh} (or U-duality related backgrounds, see, e.g. \ci{Sorokin:2011rr}) 
 interpolating  between   flat 10d space   and $AdS_2 \times S^2\times T^6$; 
 this may be  viewed as a  generalised  Bertotti-Robinson geometry. 

Below we shall  show   that  the point-like string (i.e. geodesic) motion   in this background is integrable.
We shall then   demonstrate that extended string  motion 
in the $p$-brane  geometry  with $p=0,..,6$   and in the   backgrounds  (ii), (iii)   is not integrable for generic
 values  of $Q$, 
despite integrability    being present   in the limits $Q=0$ and $Q=\infty$.

\def \const {{\rm const} }
  
\subsection{\label{sec:com}Complete integrability of geodesic motion}

The symmetries of the  metric  \eqref{Dplike} are shifts in the $x^\mu$ coordinates giving $p+1$ conserved quantities and spherical symmetry which gives $k/2$ conserved  commuting
angular momenta for even $k$ and $(k+1)/2$ for odd $k$ (generators of the Cartan subgroup of $\SO(k+1)$).
 Thus for $k\ge 3$ the spherical symmetry does not, a priori, provide us with sufficiently many conserved quantities. \\
\\
Parametrising the $k$-sphere by $k+1$ embedding  coordinates $y^i$ the effective Lagrangian for point-like string motion  is 
\begin{align}
\lagr &= f^{-2}(r)\dot{x}^\mu\dot{x}^\nu\eta_{\mu\nu}   +    f^2(r)(\dot{r}^2+r^2\dot{\vec{y}}^2)+\Lambda(\vec{y}^2-1).
\end{align}
The first integrals for $x^\mu$ and $r$ are $E^\mu=f^{-2}(r)\dot{x}^\mu$ and the Hamiltonian respectively. The equation of motion for $y^i$ reads
\be
 \partial_\tau(f^2(r)r^2\dot{y}^i)+ \Lambda y^i =0   \label{eomDplikeBraneBackground} \ .
\ee
Taking the scalar product  of this   equation with $ \vec{y}$   and with $\dot{\vec{y}}$ and using $y^iy^i=1, \  y^i\dot y^i=0$ we conclude that 
\be
\Lambda=  f^2(r)r^2\dot{\vec{y}}^2   =  f^{-2}(r)r^{-2 }  m^2   \ , \ \ \ \ \ \ \ \      m^2=  f^4(r)r^4\dot{\vec{y}}^2 =\const \ . 
\ee
Substituting this expression into \eqref{eomDplikeBraneBackground} allows one to eliminate $\Lambda$ giving
\begin{align}
 \partial_\tau(f^2(r)r^2\dot{y}^i)+m^2f^{-2}(r)r^{-2} y^i =0 \ . 
\end{align}
Multiplying this equation  by  $  f^{2}(r)r^{2} y^i $  for  fixed  $i$   and   integrating    once we obtain the constants of motion $C^i$
(no summation over $i$) 
\begin{align}
f^4(r)r^4(\dot{y}^i)^2 + m^2(y^i)^2 = C^i \ , \ \ \ \ \ \ \ \   \sum_i C^i = 2 m^2   \ . 
\end{align} 
We can rewrite the $C^i$ in terms of phase space coordinates $(y^i,\, p^i=2f^2(r)r^2\dot{y}^i)$ as 
\begin{align}
 C^i &= \frac{1}{4}(p^i)^2+m^2(y^i)^2 \ .
\end{align}
The  integrals $C^i$  are easily  checked  to  be in involution. 
Only $k$ of the $k+1$ integrals 
$C^i$  are functionally independent 
and correspond
 to the angles on the $\Omega_k$ sphere. 
 Altogether with the $p+1$ constants for the $x^\mu$ 
 coordinates and the Hamiltonian this gives in total $k+p+2$ constants of
  motion which is the same as  the number of degrees of freedom. 

\subsection{\label{sec:pbAni}Non-integrability of string motion}

Let us now  study extended  string motion in the geometry \eqref{Dplike},\eqref{fofr}. 
We  shall  choose a particular ``pulsating string''   ansatz  for  the dependence of  the string coordinates  on the  world-sheet directions 
$(\tau,\sigma)$:  we  shall   assume that  (i)  only $x^0,\, r$  and two  angles $\phi,\,\theta$  of $S^2 \subset S^k$  
(with  $\dif\Omega_2^2 = \dif\theta^2+\sin^2\theta\,\dif\phi^2$)    are  non-constant, 
and (ii)  $x^0,r,\theta$   depend only on $\tau$   while  $\phi$   depends only on $\s$, i.e. \foot{Note that  
when only one of the  angles of $S^k$ is non-constant, 1-d truncations  are not sufficient to  establish non-integrability, see below.}
\be\la{sss}
x^0=t(\tau),\qquad r=r(\tau),\qquad\phi=\phi(\sigma),\qquad\theta=\theta(\tau) \ . 
\ee
This   ansatz is consistent   with the conformal-gauge   string equations of motion   and the 
 Virasoro constraint \eqref{YdotSigmaVirasoro}. 
 The remaining string equations of motion and the  Virasoro constraint  give 
\begin{align}
\dot{t}                            &= E f^2 \label{eom1} \ , \qquad\qquad\ E=\const \ , \\
\acute{\phi}                       &= \nu =\const \ , \qquad \phi= \nu \sigma \ , \\
2\partial_\tau(f^2\dot{r})        &= \partial_r(f^2)E^2+ \partial_r(f^2)\dot{r}^2+\partial_r(r^2f^2)(\dot{\theta}^2- \nu^2 \sin^2\theta)\ ,  \\
\partial_\tau(f^2r^2\dot{\theta}) &= -f^2r^2\nu^2 \sin\theta\cos\theta\ ,  \\
E^2                                &= \dot{r}^2+r^2\dot{\theta}^2+\nu^2r^2\sin^2\theta\ .\label{viras1}
\end{align}
Thus   the string  is wrapped (with winding number $\nu$) on a  circle 
of $S^2$  whose  position in $r$ and $\theta$ changes   with time. 
The equations for $r$ and $\theta$   can be derived  from the following  effective Lagrangian
\begin{align}
\lagr &= f^2\Big( \dot{r}^2+ r^2\dot{\theta}^2-\nu^2 r^2\sin^2\theta+E^2 \Big) \ , 
\end{align}
with   the corresponding   Hamiltonian   restricted to be zero by \rf{viras1}. 
We shall    show that  this 1d 
 Hamiltonian system  is {\it not}
  integrable,  implying non-integrability of string motion in the  $p$-brane  background \eqref{Dplike}. 

Let us    choose  as an  invariant plane  $\{(r,\theta; p_r, p_\theta):\ \theta = { \pi\ov 2}, \, p_\theta=0 \}$, 
 corresponding  to the a string wrapped on 
 the equator   of $S^2$ and moving  only in $r$.
 Then   \rf{viras1}    gives $\dot{r}^2+\nu^2r^2=E^2$   which is  readily solved by (assuming  e.g. that $r(0)=0$)  
 \be\label{so}
r=\bar{r}(\tau)  = \frac{E}{\nu}\sin(\nu\tau) \ .
\ee
According to the  general method described in section 2, 
 we have to show that  small fluctuations near this special solution  are not integrable, or,  more precisely, that 
  the variational equation normal to this  surface   of solutions  parametrized by 
$E$ and $\nu$ has no Liouvillian solutions  along the curves inside the surface.

For the invariant  subspace 
$\{\theta= {\pi \ov 2}, \, p_\theta=0,\, r=\bar{r},\, p_r = 2f^2(\bar{r})\dot{\bar r}\}$ the Hamiltonian vector field is
\begin{align}
X^r = \dot{\bar{r}},\qquad
X^{p_r} = \dot{p}_r = 4ff'\dot{\bar{r}}^2+2f^2\ddot{\bar{r}},\qquad
X^\theta = 0,\qquad
X^{p_\theta} = 0
\end{align}
and thus the normal direction to this plane is along $\theta$ and $p_\theta$. 
Expanding the Hamiltonian equation for $\theta$ around $\theta= { \pi \ov 2}  ,\ \dot{\theta}=0,\ r=\bar{r}$ one obtains the NVE
($ \delta\theta\equiv \var$)
\begin{align}
\ddot{\var}+2\nu\cot(\nu\tau)\Big(1+\frac{E}{\nu}\sin(\nu\tau)\frac{f'(\frac{E}{\nu}\sin(\nu\tau))}{f(\frac{E}{\nu}\sin(\nu\tau))}\Big)\dot{\var}-
\nu^2\var &= 0 \ . \label{BRnve}
\end{align}
If  $f'/f$ is a rational function (as is the case for \rf{fofr}) 
 this equation can be algebrized through the change of variable  $\tau \to x=\sin(\nu\tau)$ giving
\begin{align}
\var''+\Big(\frac{2}{x}-\frac{x}{1-x^2}+2\frac{E}{\nu}\frac{f'(\frac{E}{\nu}x)}{f(\frac{E}{\nu}x)}\Big)\var'-\frac{1}{1-x^2}\var &= 0.
\end{align}
 Using the explicit form of $f(r)$ in  \rf{fofr} this  NVE reduces to
\begin{align}\la{vv}
\var''+\Big(\frac{2}{x}-\frac{x}{1-x^2}-2\frac{mn \nu^n \bC }{x\big( \nu^n \bC +
 E^n x^n\big)}\Big)\var'-\frac{1}{1-x^2}\var &= 0.
\end{align}
We shall assume that $\nu\not=0$ as otherwise   the string is point-like and we go back to the case of geodesic motion. 

Bringing  \rf{vv}  to the normal form by changing the variable to $\varn(x) = g(x)\var(x)$, where $g(x)$ is a suitably chosen function, one obtains
\be\la{vv1} 
\varn''+\Big[ \frac{2+x^2}{4(x^2-1)^2}-\frac{(m-1)mn^2}{x^2(1+Bx^n)^2}+\frac{mn(n-1-(n-2)x^2)}{x^2(x^2-1)(1+Bx^n)}\Big]\, \varn=0\ , 
\ \ \ \ \ \  \ \ \ \  B\equiv {E^n \ov Q \nu^n}   \ .  \ee

There are   two special limits of \rf{vv} and \rf{vv1}
\begin{align}
\bC\rightarrow 0&:\quad \varn''+\frac{2+x^2}{4(x^2-1)^2}\varn=0\ , \\
\bC\rightarrow\infty&:\quad \varn''+\frac{2x^2+x^4-4(mn)^2(x^2-1)^2+4mn(1-3x^2+2x^4)}{4x^2(x^2-1)^2}\varn=0\ , 
\end{align}
corresponding to the flat space-time and   $AdS_2\times S^2$  for $nm=1$.

We can  now apply the Kovacic  algorithm\footnote{We use Maple's function kovacicsols.} to
 determine whether
  these NVEs  do not admit Liouvillian solutions
   and thus the identity component of their Galois group is not solvable which would imply non-integrability.
   
For the special   cases $\bC\rightarrow 0$ and for $\bC\rightarrow\infty$  with $nm=1$  
  Liouvillian solutions are found 
  which is consistent with the fact that string motion  in  flat space-time and in $AdS_2\times S^2$   is  integrable.
However, for a finite value of $Q$ Liouvillian solutions do not exist for generic values of $E, \nu$
 for the cases of a  $p$-brane background with $p=0,..,6$ and  the intersecting brane 
 backgrounds of (ii) and (iii)  mentioned  above. 
 This implies non-integrability
 of string motion in those special cases of the general background \rf{Dplike},\rf{fofr}.

Finally, let us    comment on the special case of a 7-brane 
in 10 dimensions  when the transverse space is 2-dimensional, i.e. 
the string motion is described by 
\begin{align}
\lagr &= f^{-2}(r)\partial_ax^\mu\partial^ax^\nu\eta_{\mu\nu}+f^2(r)\big( \partial_ar\partial^a r+ r^2\partial_a\theta\partial^a \theta\big) \ .
\end{align}
It is easy to see that if we truncate this model to a 1-parameter system by taking each of the string 
 coordinates to be a  function of $\tau$ or $\sigma$ only  then 
 the  corresponding equations  always  admit  constants of motion for $x^\mu$ and $\theta$, 
and the Virasoro   condition then provides the solution for $r$. 
To  test  integrability  in this case 
one is to consider  some more non-trivial   truncations.
Assuming  the spatial  coordinates $x^i$ are constant  the corresponding 
effective metric is 3-dimensional:  
$ds^2 = - f^{-2}(r)  dt^2   + f^2(r) ( dr^2+ r^2 d \theta^2)$. 
 Integrability  of string motion in such a background deserves further study.

\def \ddp {{\partial_+}}   \def \ddm {{\partial_-}}

\section{\label{sec:ns5f1}String motion in  NS5-F1 background}

Let us now   include the possibility   of a non-zero $B_{MN}$ coupling  and consider 
the case   of string motion 
in  a  background produced by   fundamental strings    delocalised inside   NS5 branes \cite{fs,ns,Cvetic:1995bj,as} 
(here we use the notation $x^0=t, \ x^1= z$  and $i,j=2,...,5$) 
\begin{align}
\dif s^2 &= H_1^{-1}(r) (-\dif t^2+\dif z^2)+\dif x^i\dif x_i +H_5(r)(\dif r^2+r^2\dif\Omega_3^2)\ , 
\la{bak}\\
\dif\Omega_3^2 &= \dif\theta^2+\sin^2\theta\,\dif\varphi^2+\cos^2\theta\,\dif\phi^2\ , \no \\
B &= - H_1^{-1}(r)\,\dif t\wedge\dif z   +  \bC_5\sin^2\theta\,\, \dif\varphi\wedge \dif\phi \ , \\
H_5&= 1+\frac{\bC_5}{r^2},\qquad H_1 = 1+\frac{\bC_1}{r^2}  \ , 
\end{align}
with   the dilation 
 given  by $ e^{-2\Phi} = \frac{H_1(r)}{H_5(r)}$. 
The corresponding   string  Lagrangian in \rf{gb} 
${\cal L} = (G_{MN}   +   B_{MN}) \ddp  Y^M \ddm  Y^N$  
(where $\partial_\pm = \partial_0 \pm \partial_1$)  is 
\begin{align}
&{\cal L} =  { r^2 \ov r^2 + Q_1} \,  \ddp u\,  \ddm v   + \ddp x^i \ddm x_i   +   { r^2 + Q_5 \ov r^2} \,   \ddp r \ddm r       \no \\
 &  + ( r^2 + Q_5)  \big(   \ddp \theta  \ddm \theta     + \sin^2\theta\, \ddp \varphi  \ddm \varphi  
    + \cos^2\theta\, \ddp \phi  \ddm \phi \big)   +   Q_5 \sin^2\theta\,  (\ddp \varphi  \ddm \phi  -  \ddm \varphi  \ddp \phi )
        \ ,  \la{nsf} \\
        & \qquad \qquad \qquad     u\equiv    -t + z \ ,\ \  \qquad v \equiv  t + z   \ .  \no 
\end{align}
It    interpolates between the 
flat space  model  for $Q_1,Q_5=0$ and    the  $SL(2) \times SU(2)$    WZW model (plus 4 free directions)    for 
$Q_1,Q_5 \to \infty$.  Both of these  limits  are  obviously  integrable.

Another   integrable special case  is 
  $Q_1=0,\, Q_5\to \infty$  when we get the $SU(2)$  WZW   model plus flat directions. 
  
  The opposite limit of 
  $Q_5=0,\, Q_1\to \infty$   is described (after a rescaling of coordinates) 
  by $ {\cal L} =   { r^2} \,  \ddp u\, \ddm v     +     \ddp r \ddm r   +r^2(\ddp \theta  \ddm \theta     + \sin^2\theta\, \ddp \varphi  \ddm \varphi  
    + \cos^2\theta\, \ddp \phi  \ddm \phi)+ $ free directions.
  Solving the  equations    for $u,v$ as    ($\sigma^\pm = \tau \pm \sigma$) 
  \be r^2   \ddp u = \tilde f(\sigma^+) \ , \ \ \ \ \ \ \ \ \ \ \ \ \ \ r^2   \ddm v = f(\sigma^-)\ ,    \la{eqa} \ee
    one finds the    effective Lagrangian for $r$  being 
   \be 
    {\cal L} =      \ddp r \ddm r   -  {  f(\sigma^-) \tilde f(\sigma^+) \ov r^2} +r^2(\ddp \theta  \ddm \theta     + \sin^2\theta\, \ddp \varphi  \ddm \varphi  
    + \cos^2\theta\, \ddp \phi  \ddm \phi)\ .
    \la{rq}\ee
    As we shall find   below, this model is not  integrable.\foot{Let   us note that the constant  term in the harmonic function of the fundamental string background 
    (1 in $H_1$)   can be changed   by a combination of T-duality,  coordinate shift and another T-duality \ci{fs1,Boonstra:1997dy}, 
    so that  the fundamental string   background is not integrable for any value of $Q_1$ if it is not  integrable for large $Q_1$. 
    A similar    shift in the harmonic   function can also be  achieved  for D-brane  backgrounds  but this  involves S-duality 
    (mapping e.g. F1 to D1)  which is 
    not a symmetry of the world-sheet string action; thus  there is no contradiction with integrability of the D3-brane background at large $Q$.}

Let us note that in the limit $Q_5 \to \infty$  the action \rf{nsf}  reduces to a combination of the $SU(2)$  WZW   model,  flat directions
and the following   3-dimensional sigma model 
\be \la{lio}
{\cal L} =  {  1  \ov 1  + Q_1 e^{-2\rho} } \,  \ddp u\, \ddm v     +    \,   Q_5   \ddp \rho  \ddm \rho \ , \ \ \ \ \ \ \ \ \ \ \ \   \rho \equiv \ln r  \ ,
\ee
This model is T-dual to a  pp-wave  model with an exponential  potential function \ci{fs1,wp}  and is thus   related 
    to 
the $SL(2)$ WZW model by a combination of T-dualities  and a coordinate transformation 
(it can be interpreted, upon a rescaling of $u,v$,  as an exactly marginal deformation of the $SL(2,R)$ WZW model \ci{hs2}). 
 It should thus be integrable.  Indeed, 
solving the equations  for $z\pm t$ as in \rf{eqa}  we get the following  effective Lagrangian for $\rho$ (cf. \rf{rq}) 
    \be 
    {\cal L} =     Q_5  \ddp \rho  \ddm \rho    -   f(\sigma^-) \tilde f(\sigma^+)  (1 +   Q_1 e^{-2 \rho})   \ .
    \la{roq}\ee
Since $ f(\sigma^-) \tilde f(\sigma^+)  $   can be made  constant   by conformal redefinitions 
of  $\sigma^\pm$ (reflecting   residual gauge freedom in the conformal gauge) 
this model  is equivalent to the Liouville  theory, i.e. it is  integrable. 
Indeed, the results of our  analysis   below   are  consistent with this  conclusion. 

\

A  point-like string  does not couple to $B$-field, so geodesic motion in the  background \rf{bak}
  can be   shown to be 
integrable in the   same way as in the previous section.
To study extended string  motion let us consider the following  
ansatz describing the probe string  being  stretched along the fundamental  string 
 direction  $z$  (that may be  assumed to be  compactified 
to a circle)  and the $S^3$   angle $\varphi$, i.e. 
\be
 t = t(\tau),\quad z=z(\sigma),\quad x^i = 0,\quad r = r(\tau),\quad \theta = \theta(\tau),
 \quad  \varphi = \varphi(\sigma),\quad\phi = \phi(\tau) \ . \la{ann}
 \ee
 Then the string equations are solved by 
 \be \dot{t}=\kappa_1H_1(r(\tau))  + \kappa_2 \ , \ \ \ \ \    z= \kappa_2 \sigma \ , \ \ \ \ \ 
  \ \ \ \ \ \ 
  \varphi = \nu \sigma ,\ \ \ \ \ \ 
 \dot  \phi =  { \nu \bC_5 \ov  r^{2}(\tau)H_5(r(\tau)) } \ . \la{zz}
\ee
The resulting 1d   subsystem of equations for $r$ and $\theta$   is described by the  following effective Lagrangian
\be\la{ha}
\lagr = H_5(r)\Big[\dot{r}^2+r^2\dot{\theta}^2-\nu^2r^2\sin^2\theta-\bC_5^2\nu^2r^{-2}H_5^{-2}(r) \cos^2\theta\Big]+ \kappa_1^2 H_1(r) \ , 
\ee
which should be   supplemented by the  consequence of the Virasoro  constraint 
\be
H_5(r)\Big[ \dot{r}^2+r^2\dot{\theta}^2+\nu^2r^2\sin^2\theta + \bC_5^2\nu^2r^{-2}H_5^{-2}(r)\cos^2\theta\Big]  - \kappa_1^2H_1(r) =
 2\kappa_1\kappa_2 \ , 
\ee
implying that the Hamiltonian  corresponding to \rf{ha} is equal to $2\kappa_1\kappa_2$. This system admits the special  solution 
\be 
\theta ={\pi\ov 2} \ , \ \ \ \ \ \ \ \ \ \ 
 r=\bar{r}(\tau)\ , \ \ \ \ \ \ \   
H_5(\bar{r})(\dot{\bar{r}}^2+\nu^2\bar{r}^2) = 2\kappa_1\kappa_2+\kappa_1^2 H_1(\bar{r}) \ , 
\ee
 which may be chosen as an invariant  plane  in the phase space. 
The corresponding  Hamiltonian vector field is
\begin{align}
X^r = \dot{\bar{r}}, \qquad X^{p_r} = 2\partial_\tau(\dot{\bar{r}}H_5(\bar{r})),\qquad  X^\theta = 0, \qquad X^{p_\theta} = 0\ . 
\end{align}
Expanding along the normal direction to this  solution plane we  find  for the  NVE ($\theta = {\pi\ov 2}+\var$)
\begin{align}
 \ddot{\var} + \Big(\frac{H_5'(\bar{r})}{H_5(\bar{r})}+\frac{2}{\bar{r}}\Big)\Big(\kappa_1^2\frac{H_1(\bar{r})}{H_5(\bar{r})}+\frac{2\kappa_1\kappa_2}{H_5(\bar{r})}-\nu^2\bar{r}^2\Big)^{1/2}\dot{\var}-\nu^2\Big(1-{\bC_5^2\ov \bar{r}^{4}H_5^{2}(\bar{r})}\Big)\var &= 0.
\end{align}
Changing the  variable  $\tau$  to $x = \bar{r}(\tau)$  we  obtain the NVE in  the normal form ($'\equiv { \dif\ov \dif x}$)
\be
\varn''(x)+ U(x) \varn(x) = 0   \ , \
\ee
where 
 \begin{align}
 U(x)=  &\frac{1}{4x^2}\Big[\frac{\bC_5(\bC_5-2x^2)}{(\bC_5+x^2)^2}+\frac{4\bC_5x^4-2x^6+2\bC_1q^2(2\bC_5+x^2)}{(\bC_5+x^2)(-\bC_1q^2+
 x^2[\bC_5+x^2-q^2(1+\alpha)])}\notag \\
& \ \ \ \   +\frac{3(\bC_1q^2+x^4)^2}{(\bC_1q^2+x^2[-\bC_5-x^2+q^2(1+\alpha)])^2}\Big] = 0 \  , \ \ \ \ \ \ \ \ \ \ \ \ 
q \equiv { \kappa_1\ov \nu} \ , \ \ \ \   \alpha \equiv 2\frac{\kappa_2}{\kappa_1}\ . 
\end{align}
One finds that already for 
$\alpha=0$, i.e.  for  $\kappa_2=0$, 
 this NVE has no Liouvillian solutions for general values of $\bC_1$ and  $\bC_5$,   
  implying non-integrability of string motion  in the NS5-F1 background    with generic values of the charges. 
  Our results are summarised in the table   below.

Let us discuss some special cases. 
 Taking $\bC_1=0$ or $\bC_5=0$ one   can  absorb the remaining brane charge by rescaling $x$, $q$ and $\kappa_1$. 
 The resulting equations do not admit Liouvillian solutions for arbitrary values of $q$ which
  implies non-integrability of  string motion in the NS5-brane  or in the fundamental string backgrounds.

In the  limit  $\bC_1\rightarrow\infty$, $\bC_5\rightarrow\infty$,   in which the non-trivial part of the string action  becomes that 
of the $SL(2,R) \times SU(2)$  WZW model,  one   finds  Liouvillian solutions in agreement with the expected integrability.  
The same  applies to the case of $Q_1\to 0, \, Q_5 \to \infty$   described by the $SU(2)$  WZW model  plus free fields
and, in fact,  to  the model   with $Q_5 \to \infty$   and  arbitrary $Q_1$     described by  \rf{lio}\rf{roq}.

In the   opposite case of   $Q_5\to 0, \, Q_1 \to \infty$ described by \rf{rq} integrability appears to be absent 
as we did not find Liouvillian solutions  (see {table}).\foot{Let us note  that the  results about  the existence of Liouvillian solutions turn out  to be 
  independent of whether the truncated solutions contains any contribution from the B-flux of the 
  fundamental string solution ($\alpha\ne 0$ or $\alpha=0$), reflecting a special choice of  our ansatz \rf{ann}.}

 Since  2d duality transformations  of string coordinates  preserve  (non)integrability,    similar conclusions can be reached for
 string backgrounds related to  this  NS5-F1   background  or to its limits    via T-dualities (and coordinate transformations).
 In particular,   this applies to the  pp-wave background related  to the fundamental string   by   T-duality  \ci{fs1}  which is  studied in 
    in Appendix  \ref{sec:ppWave}. 

\def \g {\gamma} 

\begin{figure}[ht]
\renewcommand{\arraystretch}{1.5}
\begin{tabular}{|ll|>{\centering\arraybackslash}m{9.1cm}|>{\centering\arraybackslash}m{2.7cm}|}
\hline
\multicolumn{2}{|c|}{limit} & $U(x)$ in  NVE \ $\varn''(x)+U(x)\varn(x)=0$,\ \ \ $\g\equiv { \bC_5\ov \bC_1}$ 
& rescaling\\ 
\hline
$\bC_5\rightarrow 0$     & $\bC_1\rightarrow 0$ & $\frac{2q^2(1+\alpha)+x^2}{4(q^2(1+\alpha)-x^2)^2}$ & - \\
$\bC_5\rightarrow\infty$ & $\bC_1\rightarrow\infty$ & $\frac{(\g-q^2\alpha)(\g x^2+q^2(2-x^2\alpha)}{4(\g x^2-q^2(1+x^2\alpha))^2}$ & $\alpha\rightarrow \bC_1\alpha$\\
$\bC_5\rightarrow\infty$ & $\bC_1\rightarrow 0$ & $\frac{1}{4x^2}$ & -\\
$\bC_5\rightarrow 0$     & $\bC_1\rightarrow\infty$ & $\frac{\bC_1^2q^4+10\bC_1q^2x^4+x^8-2\bC_1q^4x^2\alpha+2q^2x^6\alpha}{4(\bC_1q^2x-x^5+q^2x^3\alpha)^2}$ & -\\
$\bC_5\rightarrow\infty$ & $\bC_1 \ne 0$ & $\frac{(\g -q^2(1+\alpha))(\g x^2-q^2(-2+x^2(1+\alpha)))}{4(\g x^2-q^2(1+x^2(1+\alpha)))^2}$ & $x\rightarrow \sqrt{\bC_1}x$, $\kappa_{1,2}\rightarrow\sqrt{\bC_1}\kappa_{1,2}$ \\
%
$\bC_5 \ne 0$            & $\bC_1\rightarrow\infty$ & \vskip-2em{\scriptsize\begin{align*}\Big[ 
\g^{-2}q^4(x^2-2)+2\g^{-1}q^2(1+x^2)^2(1+5x^2)-2\g^{-1} q^4(1+5x^2+x^4)\alpha\\   +\ x^2((1+x^2)^2-q^2\alpha)((1+x^2)^2+q^2(2x^2-1)\alpha)\Big]\\ \times \Big[ 4(1+x^2)^2(x^2+x^4-q^2(\g^{-1}+x^2\alpha))^2\Big]^{-1}\end{align*}}\vskip-3em\,
& $x\rightarrow \sqrt{\bC_5}x$, $\kappa_{1,2}\rightarrow\sqrt{\bC_5}\kappa_{1,2}$\\
$\bC_5\rightarrow 0$     & $\bC_1 \ne 0$ & $\frac{x^8+q^4(1-2x^2(1+\alpha))+2q^2x^4(5+x^2(1+\alpha))}{4x^2(x^4-q^2(1+x^2(1+\alpha)))^2}$ & $x\rightarrow \sqrt{\bC_1}x$, $\kappa_{1,2}\rightarrow\sqrt{\bC_1}\kappa_{1,2}$\\
$\bC_5 \ne 0$            & $\bC_1\rightarrow 0$ & $\frac{(1+x^2)^4+2q^2(x^2-1)(1+x^2)^2(1+\alpha)-q^4(2x^2-1)(1+\alpha)^2}{4x^2(1+x^2)^2(1+x^2-q^2(1+\alpha))^2}$ & $x\rightarrow \sqrt{\bC_5}x$, $\kappa_{1,2}\rightarrow\sqrt{\bC_5}\kappa_{1,2}$\\
\hline
\end{tabular}
\renewcommand{\arraystretch}{1.0}
 \begin{tabular}{|ll|>{\centering\arraybackslash}m{2.5cm}
 |>{\centering\arraybackslash}m{2.0cm}|>{\centering\arraybackslash}m{2.0cm}|>{\centering\arraybackslash}m{2.0cm}|>{\centering\arraybackslash}m{2.05cm}|}
\hline
\multicolumn{2}{|c|}{\multirow{2}{*}{limit}} & \multirow{2}{*}{spacetime} & \multicolumn{2}{|>{\centering\arraybackslash}m{4.3cm}|}{2-form contributes to the truncated system} & \multicolumn{2}{|>{\centering\arraybackslash}m{4.3cm}|}{NVE has Liouvillian solutions}\\ \cline{4-7}
 &  & &  $\alpha=0$ & $\alpha\ne 0$ & $\alpha=0$ & $\alpha\ne 0$\\
\hline
$\bC_5\rightarrow 0$ & $\bC_1\rightarrow 0$ & $\R^{1,9}$ & \xmark & \xmark &\cmark & \cmark\\
$\bC_5\rightarrow\infty$ & $\bC_1\rightarrow\infty$ & $AdS_3\times S^3\times \R^4 $ & \cmark & \cmark & \cmark & \cmark\\
$\bC_5\rightarrow\infty$ & $\bC_1\rightarrow 0$ &  $\R^{1,6}\times  S^3$  & \cmark & \cmark & \cmark & \cmark\\
$\bC_5\rightarrow 0$ & $\bC_1\rightarrow\infty$ & -        &  \xmark & \cmark & \xmark &\xmark\\
$\bC_5\rightarrow\infty$ & $\bC_1 \ne 0$ & - & \cmark & \cmark & \cmark & \cmark\\
$\bC_5 \ne 0$ & $\bC_1\rightarrow\infty$ & - & \cmark & \cmark & \xmark & \xmark\\
$\bC_5\rightarrow 0$ & $\bC_1 \ne 0$ & F1  & \xmark & \cmark & \xmark & \xmark\\
$\bC_5 \ne 0$ & $\bC_1\rightarrow 0$ & NS5  & \cmark & \cmark & \xmark & \xmark\\
\hline
\end{tabular}
\label{resultsTable1}
\end{figure}
\section{Concluding remarks}

We have shown that for various $p$-brane backgrounds, for which the   string sigma model 
interpolates between integrable flat and coset  or  WZW models and  which has integrable   geodesics,  
 the corresponding  extended classical string motion is not integrable in general.

To demonstrate  non-integrability we considered   particular extended string motion  for which
the dynamics reduces to an effective one-dimensional system. The latter has  special integrable subclasses of solutions,
but perturbing near them  leads to linear second-order  differential equations 
that cannot be solved in quadratures. 

It would   be interesting to understand better     why   switching on a
non-zero D3-brane   charge  or  moving away from the ``throat'' ($AdS_5 \times S^5$) region  leads to a breakdown of string integrability. 

Together with similar previous results  \ci{za,pan,*pan1,Basu:2012ae}  this  supports  the   expectation   that     integrability   of  classical string motion 
is  a rare   phenomenon. String integrability is thus a much more restrictive constraint than {integrability of particle motion}.
Still,  in the absence of a  general classification of integrable   2d   sigma models   one  cannot rule out 
the possibility that there are  still interesting examples of integrable backgrounds
(that  cannot be obtained from known solvable cases by  coordinate transformations combined with T-dualities, cf. \ci{Tseytlin:1995fh,Russo:1994cv}), 
  that remain to be discovered.


\section*{Acknowledgments}
AAT  is grateful to G. Gibbons for a useful comment. 
He also   thanks K. Sfetsos for pointing out possible connection to ref. \ci{Prezas:2008ua}. 
This  work was  supported by the    STFC grant ST/J000353/1.   
A.A.T. was  also  supported by the ERC Advanced  grant No.290456.


\appendix
\section{\label{sec:dgt}Integrability and differential Galois theory}

Integrability in the sense of Liouville implies that the equations of motion are solvable in
 quadratures, i.e. the solutions are combinations of integrals of rational functions, exponentials, logarithms and 
 algebraic expressions.
  Solutions of this type are referred to as Liouvillian functions. 
  Thus the question of integrability is related to the question of when differential equations admit only Liouvillian solutions. 
  This is answered by differential Galois theory and the main  integrability theorem, which our 
  analysis relies on, can be stated as follows \cite{2011arXiv1104.0312A}:

{\bf Theorem} \ \ {\it Let $H$ be a Hamiltonian defined on a phase space manifold $M$ with an associated Hamiltonian vector field $X$. Let $P$ be a submanifold of $M$ which is invariant under the flow of $X$ with the invariant curves $\Gamma$ and let $X|_P$ be completely integrable.
Also let $G$ be the differential Galois group of the variational equation of the flow of $X|_P$ along $\Gamma$ normal to $P$. 
If $X$ is completely integrable on $M$ then the largest connected algebraic subgroup of $G$ which contains the identity is abelian. } 

Since  differential Galois theory is not frequently used in physics we shall give a brief introduction and illustrate  some basic concepts 
 on a simple example below. We  shall follow  ref.\cite{Magid94lectureson}. First, let us give some basic definitions:
\begin{enumerate}
\item A differential field is a field $F$ with a derivation $D_F$, which is an additive map $D_F: F\rightarrow F$ thus satisfying 
 $D_F(ab)=D_F(a)b+aD_F(b)\,,\ \forall  a,b\in F$.
\item A differential homomorphism/isomorphism between two differential fields $F_1$ and $F_2$ is a homomorphism/isomorphism $f:\ F_1 \rightarrow F_2 $ which satisfies $D_{F_2} (f(a)) = f(D_{F_1} (a))\, ,\ \forall a\in F_1$.
\item A linear differential operator $L$ is  defined by 
\begin{align}
L  (y) &= y^{(n)}+a_{n-1}y^{(n-1)}+...+a_0 y
\end{align}
where $y^{(n)} = D^n_F(y)$ and $a_i\in F$.
\item A differential field $E$ with derivation $D_E$ is called a differential field extension of a differential field $F$ with derivation $D_F$ iff $E\supseteq F$ and the restriction of $D_E$ to $F$ coincides with $D_F$.
\item Elements $a\in F$ whose derivative vanishes are called constants of $F$. The subfield of constants of $F$ is denoted by $C_F$.
\end{enumerate}
 In order to investigate the types of solutions a differential equation admits one has to encode the relations and symmetries of the independent solutions. Relations can be encoded in the Picard-Vessiot  extension of a differential field which includes the solutions of $L(y)=0$:
\begin{definition}
An extension $E\supseteq F$ is called a Picard-Vessiot extension of $F$ for $L(y)=0$ iff:
 
 $E$ is generated over $F$ as a differential field by the set of solutions of $L(y)=0$ in $E$; 
 
 $E$ and $F$ have the same constants;
 
 $L(y)=0$ has $n$ solutions in $E$ linearly independent over the constants.

\end{definition}
One can show that the Picard-Vessiot extension always exists for differential fields with an algebraically closed field of constants and  this extension is
 unique up to isomorphisms.

Symmetries of the solution space are linear transformations which,  applied to 
 a solution,  give a new solution. The
   simplest formulation is in terms of automorphisms of the Picard-Vessiot extension and leads to the notion of the differential Galois group:
\begin{definition}
For a differential field $F$ with the Picard-Vessiot extension $E\supseteq F$ the differential Galois group $G(E/F)$ is the group of differential automorphisms $\sigma: E\rightarrow E$ whose restriction to $F$ is the identity map, i.e.
\begin{align}
G(E/F) &= \left\{\sigma: E\rightarrow E\left|\,\sigma(a)=a\,\ \ \forall a\in F\right.\right\}. \nonumber
\end{align}
\end{definition}
 A solution $y(x)$  of a linear differential equation $L(y)=0$ is called \

 {\it algebraic} if it is a root of a polynomial over $F$;
 
 {\it primitive} if $y'(x)\in F$, i.e. $y(x)=\int^x f(z)\,\dif z$ for $f(x)\in F$;
 
 {\it exponential} if ${y'(x)\ov y(x)}\in F$, i.e. $y(x)=\exp\left(\int^x f(z)\,\dif z\right)$   with  $f(z)\in F$. 

\noindent 
\begin{definition}
The solution  $y(x)$ is  called  {\it Liouvillian} if there exist differential extensions $E_i,\,i=0,..,m$
\begin{align}
F = E_0\subset E_1\subset ... \subset E_m = E \nonumber
\end{align}
 such  that $y(x)=y_m(x) \in E$,  $E_i = E_{i-1}(y_i(x))$  and    $y_i(x)$ is algebraic, primitive or exponential over $E_{i-1}$.
\end{definition}

We will be interested in second order homogeneous linear ordinary differential equations
\begin{align}
 y''(x)+a(x)y'(x)+b(x)y(x) &= 0\label{2ndOrderODE}\ , 
\end{align}
where $a(x), b(x)\in F$ and $F=\mathbb{C}(x)$ is the differential field of rational functions
over complex numbers.
 The equation \eqref{2ndOrderODE} can be cast into the normal form
\bea
&& y''(x) + U(x)\, y(x)=0\ , \la{a1} \\
&&
U(x)=-\frac{1}{4}a^2(x)-\frac{1}{2}a'(x)+b(x)\ ,\ \ \ \ \   U(x)\in F\label{normalFormODE}
\eea
by the transformation $y(x)\rightarrow y(x)\exp\left(-\frac{1}{2}\int^x a(z)\dif z\right)$.

 Let us identify the corresponding Galois group. Let $E$ be a Picard-Vessiot extension of $F=\mathbb{C}(x)$ and $z_1(x),z_2(x)\in E$ be two independent solutions of \rf{a1}. The Galois group $G(E/F)$ consists of differential automorphisms $\sigma$ 
 which  act on the space of solutions of \rf{a1}, i.e.  map solutions  to solutions. Thus 
\begin{align}
\sigma(z_i(x)) = C_{ij}z_j(x) \ , \ \ \ \ \ \ \ \    i,j=1,2 \ .  \la{a2}
\end{align}
Defining the Wronskian $
w(z_1(x), z_2(x)) = \det\Big(\begin{array}{cc}
z_1(x) & z_2(x)\\
z_1'(x) & z_2'(x)\end{array}\Big)
$
we obtain
\begin{align}
\sigma(w(z_1(x), z_2(x))) = \det C\  w(z_1(x), z_2(x))
\end{align}
where $ \det C  $ is the determinant of the $C_{ij}$  matrix in   \rf{a2}.
The Wronskian is a constant since $w'(x)=z_1(x)z_2''(x)-z_2(x)z_1''(x)=0$ by virtue of \eqref{a1}, i.e. 
 $w(x)\in F$. Therefore,  $\sigma(w(x)) = w(x)$, giving $\det C=1$.
 Thus  the Galois group for \eqref{a1} is a subgroup of $\SL(2,\C)$. 
 For more general equations the Galois group is a subgroup of $\GL(n,\C)$.

Let us   now  address  the  main question:  whether  a differential equation can be solved  in terms of 
 Liouvillian functions. Since the minimal differential extension generated by all solutions is the Picard-Vessiot extension we therefore require that the Picard-Vessiot extension is generated by a tower of Liouvillian extensions. By the fundamental theorem of differential Galois theory this tower of extensions is related to subgroups of the Galois group through a bijective correspondence. One can show that for the Picard-Vessiot extension to be Liouvillian, 
 the   {\it  identity-connected component }  of the Galois group $G^0$ has to be {\it solvable} 
 (see theorem 25 in \cite{Kolchin1948}).\foot{A group  is  solvable if it has a subnormal series whose factor
   groups are all abelian, that is, if there are subgroups  $ \{1\} \leq   G_1 \leq...\leq G_n=G$  such that  
   $G_{k-1} $ is normal in $G_{k}$ (i.e. invariant under conjugation) 
     and 
   $G_k/G_{k-1}$  is an abelian group for all $k=1, ..., n$.} 

%
%
%
%
A simple  example is provided by  the equation
$ y''(x)+2x y'(x) = 0  $
which has the normal form  
\begin{align}
y''(x)-(x^2+1)y(x) &= 0.
\end{align}
For this equation  $F=\C(x)$ is the field of rational functions over complex numbers. The solution space   may be represented as  
 $\C\,(e^{\frac{1}{2}x^2})\oplus \C(\,e^{\frac{1}{2}x^2}\int\, e^{-x^2})$.
 The Picard-Vessiot extension is  $E=\C(x,e^{\frac{1}{2}x^2},e^{-x^2},\int e^{-x^2})$    and the tower 
 of extensions of $E$ is 
\begin{align}\nonumber
\C(x)\subset \C\left(x,e^{\frac{1}{2}x^2}\right)\subset \C\left(x,e^{\frac{1}{2}x^2},e^{-x^2}\right)\subset E=\C\Big(x,e^{\frac{1}{2}x^2},e^{-x^2},\int e^{-x^2}\Big)\ . 
\end{align}
 Then the  Galois group $G(E/F)$  and its identity component $G^0$ are
\begin{align}
G=G^0= \Big\{\Big(
\begin{array}{cc}a & 0\\ c & 1/a\end{array}\Big)  ,\ \      
a\not=0,c \, \in\C\Big\}
\end{align}
This is a solvable group since it is a subgroup of $SL(2,\C)$ whose  algebraic subgroups are all solvable except for  $SL(2,\C)$ itself.

\section{\label{sec:ppWave} String motion in  pp-wave background}

Here we shall   study string motion in  a  pp-wave  metric ($i=1,..,d$) 
\begin{align}\la{met}
\dif s^2 &= \dif u\,\dif v+H(u,x)\,\dif u^2+\dif x^i\dif x_i  \ . 
\end{align}
In conformal gauge the  equation  for $v$ reads
$\partial_+\partial_-u = 0$
and is solved by
$u = f(\sigma^+) + \td f (\sigma^-)$.
We can fix the residual conformal symmetry by choosing  $u=p \tau$.
Then $v$ is determined from the Virasoro   constraints   and we  get  the following  effective Lagrangian for $x_i$ 
\begin{align}
\lagr &= \dot{x}^i\dot{x}^i-\acute{x}^i\acute{x}^i+p^2H(u,x),\label{pp}
\end{align}
which describes light-cone gauge    string motion in a 
potential.

One  familiar  example  is the Ricci flat space with 
$H(x) =  \mu_{ij} x^i x^j, \ \mu^i_i=0. $
Another  is the  pp-wave limit of  the $AdS_5\times S^5$   background 
for which $H(x) = x^i x_i $ \cite{Blau:2002dy}. Here the string motion  takes place 
 in  a  quadratic potential  and is obviously    integrable. 
  Penrose limits of the brane backgrounds we consider in this paper also take
  the pp-wave metric form  with $H(u,x)=h_{ij}(u)x^i x^j$  (see e.g. \ci{Fuji:2002vs,*Ryang:2002fj}).
  In the light-cone gauge string  motion  takes place in   a  quadratic potential 
  with a $\tau$-dependent  coefficient  and it is  solvable   \cite{Papadopoulos:2002bg,*Blau:2003rt} 
  but formally the resulting model  is not integrable.\foot{For the relevant case of $h \sim { 1 \ov \tau^2}$ 
  the Kovacic algorithm gives   no Liouvillian solutions   for string modes that    depend  on $\sigma$.}


  There are also    other  integrable examples   with $H(x)$   corresponding, for example,  to  the Liouville 
 or  Toda  potential.

The case that we shall  study below is the pp-wave solution   T-dual to the fundamental string \cite{fs,fs1} for which 
 $H$ is a  harmonic function 
\begin{align}
H(x) = 1+\frac{Q}{r^{d-2}} \ , \ \ \ \ \ \ \ \ \  \ \ \ \  r^2 = x^i x_i  \ . 
\end{align}
Not surprisingly, the  conclusions will be the same as for the fundamental  string in section 4: 
the geodesic   motion is   integrable but  an extended string motion is not. 

In the  point-like  string limit \rf{pp}  takes   the form 
(we set $x^i = ry^i$ with $\vec{y}^2=1$)
\begin{align}
 \lagr &= \dot{r}^2+r^2\dot{\vec{y}}^2+p^2H(r)+\Lambda(\vec{y}^2-1) \ . 
\end{align}
The corresponding  geodesic  motion is completely integrable since we have $d-1$ constants of motion from the coordinates $y^i$ 
(see section \ref{sec:com}) plus the   Hamiltonian.

Let us now consider  an extended  string  moving only in a 3-space 
$dx^i dx_i = dr^2 + r^2 ( d \theta^2 + \sin^2 \theta\, d \phi^2$)
and choose the following ansatz (similar to the one in \rf{sss})
\begin{align}
r=r(\tau),\ \ \qquad\phi=\phi(\sigma)=\nu \sigma,\ \  \qquad\theta=\theta(\tau)\ . 
\end{align}
Then  
the effective Lagrangian   for $r$ and $\theta$  takes the form
\begin{align}
\lagr = \dot{r}^2+r^2\dot{\theta}^2 -\nu^2 r^2\sin^2\theta +p^2H(r) \ . \la{lal}
\end{align}
If we also assume that $v=0$ (i.e. the  string also moves in the light-cone direction) 
then  the  Virasoro   condition   for \rf{met}   is equivalent to the vanishing of the Hamiltonian 
corresponding to \rf{lal}
\begin{align}
\dot{r}^2+r^2\dot{\theta}^2+\nu^2r^2\sin^2\theta-p^2H(r)=0 \ . \la{jo}
\end{align}
Restricting motion to   the invariant plane   $\{(r, p_r, \theta,p_\theta):\theta = {\pi\ov 2} , \,  p_{\theta}=0\}$
 gives a  one-dimensional integrable system parametrised by $\nu$ and $p$ 
 with the Hamiltonian as the constant of motion. 
 Imposing the  zero  Hamiltonian  constraint \rf{jo}  gives the {solution} for $r$
%
%
\begin{align}
\dot{r}^2 + \nu^2r^2 &= p^2H(r) \ , \ \ \ \ \ \ \  \ \ \  r= \bar r (\tau) \ . 
\end{align}
The Hamiltonian vector field on the plane of these solutions is
\begin{align}
X^r = \dot{\bar{r}},\qquad
X^{p_r} = 2\ddot{\bar{r}} = -2\nu^2\bar{r}+p^2H'(\bar{r}),\qquad
X^\theta = 0,\qquad
X^{p_\theta} = 0
\end{align}
and the direction normal to this plane is along $\theta$ and $p_\theta$. Expanding the equation of motion for $\theta$ gives the NVE
($\var = \delta\theta= \theta - {\pi\ov 2} $)
\begin{align}
\ddot{\var}+2\frac{\dot{\bar{r}}}{\bar{r}}\dot{\var}-\nu^2\var &= 0\ . 
\end{align}
 Changing the independent variable $\tau\rightarrow r=\bar{r}(\tau) $ one obtains ($ '\equiv{\dif\ov \dif r}$)
\begin{align}
\var''+\Big(\frac{2}{r}+\frac{\ha{p^2}H'(r)-\nu^2 r}{p^2H( r)-\nu^2 r^2}\Big)\var'-\frac{\nu^2}{p^2H(  r)-\nu^2  r^2}\var &= 0
\end{align}
Bringing this equation to the normal form via the change of variable  $\var(r) = g(r)\varn(r)$ we get 
\begin{align}
&\qquad \qquad \qquad \qquad  \qquad \qquad   \varn'' +   U(r)\, \varn  =0 \ , \\
& U(r) = 
\frac{-(d-2)q^4Q\big[ (d-6)Qr^2+4(d-3)r^d\big] +4q^2r^d\big[ (d^2 -2d + 2 )Qr^2+2r^d\big]
+4r^{2+2d}}{16\big[  q^2(Qr^2+r^d)-r^{2+d}\big]^2}= 0 \ ,\no
\end{align}
where $q \equiv { p\ov \nu}$. For $2<d<13$ the Kovacic algorithm does not yield Liouvillian solutions for general finite  non-zero values of $Q$
 and $q$  which implies non-integrability.

Note that  in the limit $Q\rightarrow 0 $,  when the   metric becomes flat,  we get 
\begin{align}
 \varn''+\frac{2q^2+r^2}{4(q^2-r^2)^2}\varn = 0,
\end{align}
which has  Liouvillian solutions  in agreement with   with flat space  integrability.

\parskip=0.pt
\baselineskip 13pt 

\bibliographystyle{utphysM}
\bibliography{v2_rev2}

\ed